\documentclass[showpacs,preprint,amsmath,amssymb]{revtex4}
\usepackage{graphicx}
\usepackage{dcolumn}
\usepackage{bm}
\begin{document}
\title{Giant optical anisotropy in a single InAs quantum dot in a very dilute quantum-dot ensemble}
\author{I. Favero, G. Cassabois,$^\ast$ A. Jankovic, R. Ferreira, D. Darson, C. Voisin, C. Delalande and Ph. Roussignol}
\affiliation{Laboratoire Pierre Aigrain, Ecole Normale
Sup\'erieure, 24 rue Lhomond 75231 Paris Cedex 5, France}
\author{A. Badolato and P. M. Petroff}
\affiliation{Materials Department and Electrical and Computer
Engineering Department, University of California, Santa Barbara,
California 93106, USA}
\author{J. M. G\'{e}rard}
\affiliation{CEA-CNRS-UJF "Nanophysics and Semiconductors"
Laboratory, CEA/DRFMC/SP2M, 17 rue des Martyrs 38054 Grenoble
Cedex 9, France}
\date{\today}
\begin{abstract}
We present the experimental evidence of giant optical anisotropy
in single InAs quantum dots. Polarization-resolved
photoluminescence spectroscopy reveals a linear polarization ratio
with huge fluctuations, from one quantum dot to another, in sign
and in magnitude with absolute values up to 82\%. Systematic
measurements on hundreds of quantum dots coming from two different
laboratories demonstrate that the giant optical anisotropy is an
intrinsic feature of dilute quantum-dot arrays.
\end{abstract}

\maketitle

Semiconductor quantum dots (QDs) offer a huge potential for the
development of nanodevices. Recent papers report breakthroughs
where a single QD serves as the physical support for the
realization of an all-optical quantum gate \cite{li}, a
single-photon source \cite{michler} for indistinguishable photon
generation \cite{santori}, and an optically triggered
single-electron source \cite{zrenner}. In some novel QD devices,
the polarization of the QD optical response could play a central
role on their characteristics. For example, the carrier quantum
cascade in a QD has been proposed for the generation of
polarization-entangled Bell states in quantum information
processing \cite{benson}. In the field of spintronics, the
conversion of photon polarization to electron spin has the
potential to provide spin-polarized carriers \cite{pryor}.

Assuming rotational symmetry of the QD along the [001] growth
axis, one predicts, for the fundamental interband transition, two
bright electron-hole pair states which are degenerate and excited
by orthogonal circularly polarized light states. However,
different microscopic effects break the QD rotational invariance
such as the atomistic asymmetry of the crystal \cite{zunger}, and
the anisotropy of the QD shape or composition
\cite{zunger,ivchenko}. Because of the electron-hole exchange
interaction the new QD eigenstates correspond to two linearly
polarized transitions, $\mid$$X$$>$ and $\mid$$Y$$>$, which are
aligned along two orthogonal axes of the nanostructure.
Theoretical calculations predict so far almost equal oscillator
strengths for both transitions \cite{zunger}.

Polarized photoluminescence (PL) spectroscopy in strongly confined
InAs QDs has been employed by various groups to study, in single
QDs, the fine structure of the fundamental transition and the
degree of linear/circular polarization of its emission
\cite{toda,bayer,finley,stevenson,urbaszek}. Linearly polarized
doublets are observed with an energy splitting ranging from a few
tens of $\mu$eV \cite{stevenson,urbaszek} to several hundreds of
$\mu$eV \cite{bayer,finley}. If the PL intensity of both linearly
polarized transitions is added, a net linear optical anisotropy of
a few percents is obtained (except for the value of thirty
percents mentioned in Ref.~\cite{santoriB}), in agreement with the
available theoretical framework.

In this Letter, we present the experimental evidence of giant
optical anisotropy in single InAs QDs. Polarization-resolved PL
spectroscopy in single QDs reveals a linear polarization ratio
which fluctuates, from one dot to another, in sign and in
magnitude with absolute values up to 82\%. We do not observe any
dependence of the linear polarization on incident power,
temperature and excitation laser polarization. We interpret the QD
optical anisotropy as due to different oscillator strengths for
the two linearly polarized states. Systematic measurements on
hundreds of QDs coming from two different laboratories allow us to
demonstrate that the giant optical anisotropy is an intrinsic
feature of dilute QD arrays. On the contrary, the optical
anisotropy is much smaller for QDs belonging to dense arrays.

Single-QD spectroscopy at low temperature is performed by standard
micro-PL measurements in the far field using a microscope
objective in a confocal geometry. The excitation beam is provided
by a linearly polarized He:Ne laser. The signal is detected, after
spectral filtering by a 32 cm monochromator, either by a
LN$_{2}$-cooled charge-coupled-device (CCD) for high
signal-to-noise measurements with a spectral resolution of 180
$\mu$eV, or by a low noise Si-based photon counting module for
ultra-precise linewidth measurements using interferometric
correlation with a sub-$\mu$eV spectral resolution
\cite{kammerer}. Polarization-resolved experiments are implemented
by using a rotating half-wave retarder and a fixed linear
polarizer in front of the spectrometer in order to avoid detection
artifacts due to the anisotropic response function of the setup.
The analysis axes $X$ and $Y$ are orthogonal and the $X$ axis is
parallel either to the [110] or to the [1$\overline{1}$0]
crystallographic direction of the sample.

We present experimental data obtained for samples containing a
single layer of InAs QDs grown on standard (001) GaAs substrates.
Unlike previous work devoted to the study of the average
polarization of QD ensembles \cite{yu,henini}, we explore here the
PL polarization properties in \textit{single} QDs in
\textit{dilute} and \textit{dense} QD arrays. Our self-assembled
QDs (base~$\sim$~20 nm, height~$\sim$~2 nm) are obtained by
molecular beam epitaxy in the Stranski-Krastanow growth mode. The
transition from layer-by-layer growth to three-dimensional
islanding occurs for a critical thickness of 1.7 monolayer in
InAs. Owing to a gradient of the InAs layer thickness in our
samples, three different regions can be identified on the surface
of the wafers \cite{cassabois}: region (I) in two-dimensional
growth mode with no QDs, region (III) in three-dimensional growth
mode with an areal QD density of 10$^3$$\mu$m$^{-2}$ and an
intermediate region (II), the so-called border region at the onset
of QD nucleation, corresponding to a very dilute QD ensemble. The
QDs studied in this paper come from three distinct wafers labelled
$A$, $B$ and $C$, where we stress that $C$ was grown in a
different laboratory from $A$ and $B$. In the following, we
present measurements on a set of four samples: $A_{(II)}$ and
$C_{(II)}$ corresponding to dilute QD arrays in the border region
(II) of wafers $A$ and $C$, and $B_{(III)}$ and $C_{(III)}$
corresponding to dense QD arrays in region (III) of wafers $B$ and
$C$. Let us finally note that mesa patterns \cite{marzin} are
processed in our four samples in order to perform
polarization-resolved single QD spectroscopy in similar
conditions, although the etching of mesas is not necessary in
$A_{(II)}$ and $C_{(II)}$ because of the low QD density.

\begin{figure}
\includegraphics[scale=0.73]{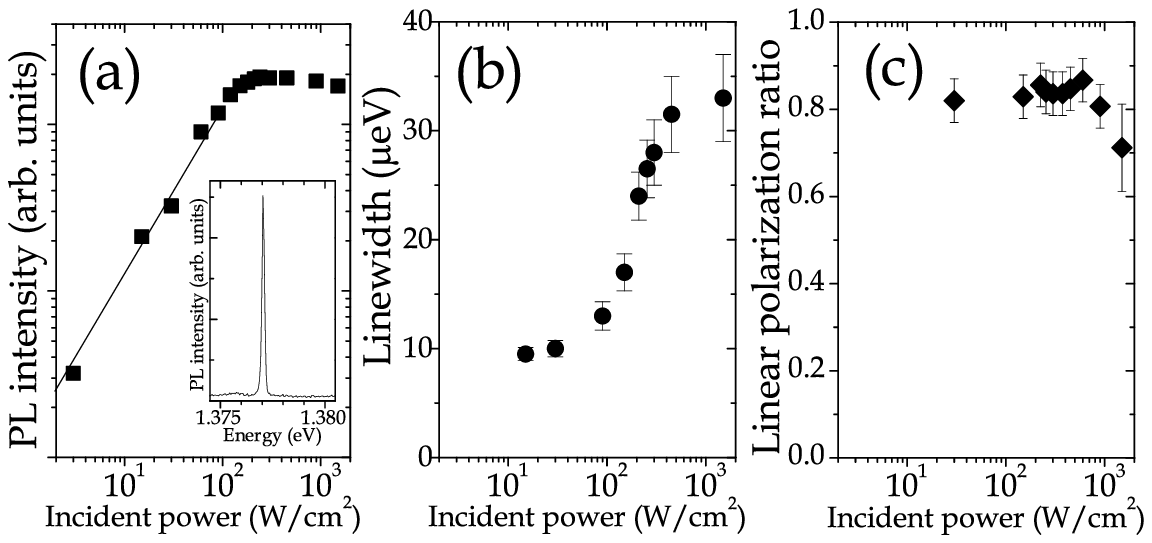}
\caption{Photoluminescence spectroscopy versus incident power for
QD1 at 10K. (a) Spectrally integrated photoluminescence intensity,
(b) linewidth (c) linear polarization ratio. (Inset):
photoluminescence spectrum with an incident power of 30 W/cm$^2$.}
\label{PowerDepend}
\end{figure}

We first present a striking experimental evidence of giant optical
anisotropy obtained for a single QD in sample $A_{(II)}$,
hereafter called QD1. A low-temperature PL spectrum recorded with
the CCD is displayed in the inset of Fig.~\ref{PowerDepend}. We
observe a resolution-limited sharp line corresponding to
electron-hole recombination in the single QD. Rotation of the
half-wave retarder produces a strong variation of the PL signal
intensity which is maximum and minimum for the $X$ and $Y$
analysis axes, respectively. The ratio $I_X$/$I_Y$ of the measured
intensities is as high as 10 for this single QD, giving a linear
polarization ratio $P_L$=($I_X$-$I_Y$)/($I_X$+$I_Y$) of
0.82$\pm$0.05. Rotation of the excitation laser polarization does
not induce any modification. Further inspection of the emission
polarization with a quarter-wave retarder allows us to exclude the
existence of two orthogonal elliptically polarized states. These
observations are in agreement with the expected linearly polarized
transitions in QDs of reduced symmetry. We are not able to resolve
the fundamental transition doublet corresponding to the two
different polarizations. The effective energy splitting resolution
of our setup is roughly 45 $\mu$eV which gives us an upper bound
for the splitting energy of the fundamental doublet.

We now discuss the variation of the linear polarization ratio with
the experimental conditions. In Fig.~\ref{PowerDepend} we display
the different power-dependent measurements performed on QD1. The
spectrally integrated intensity [Fig.~\ref{PowerDepend}(a)] shows
a linear increase with the incident power, followed by the
standard saturation behavior. Interferometric correlation
measurements of the QD linewidth allow us to evidence that the QD
line keeps a Lorentzian profile, and that its width, which is
independent from the polarization analysis axis, increases by a
factor 3 with incident power [Fig.~\ref{PowerDepend}(b)]. Such a
phenomenology reveals environment effects on the QD optical
properties, likely coming from Coulomb correlations between
carriers in the QD and its environment \cite{uskov}. When
increasing the incident power, the linear polarization ratio
remains fairly constant [Fig.~\ref{PowerDepend}(c)]. We can thus
exclude any difference in the capture efficiency for the
$\mid$$X$$>$ and $\mid$$Y$$>$ states since their populations
become equal above saturation. Temperature-dependent measurements
in QD1 (not shown here) indicate identical features as follows.
The QD transition energy follows the InAs bandgap energy when
increasing the temperature. The linewidth variation exhibits the
usual thermo-activated broadening due to phonon dephasing
\cite{kammerer} and the linewidth increases by a factor 16 between
10K and 70K. Nevertheless the linear polarization ratio is
unchanged.

Since there is no dependence on incident power, temperature and
excitation laser polarization of the linear polarization ratio, we
thus conclude that the optical anisotropy is not related to
carrier relaxation nor to environment effects. This phenomenon,
with the striking example of QD1 but also observed for all
investigated QDs, appears to be intrinsically bound to the fine
structure of the QD fundamental transition. In the picture of the
linearly polarized doublet in a QD of reduced symmetry, we
interpret the QD optical anisotropy as due to \textit{different
oscillator strengths} for the two linearly polarized states
$\mid$$X$$>$ and $\mid$$Y$$>$.

\begin{figure}
\includegraphics[scale=0.85]{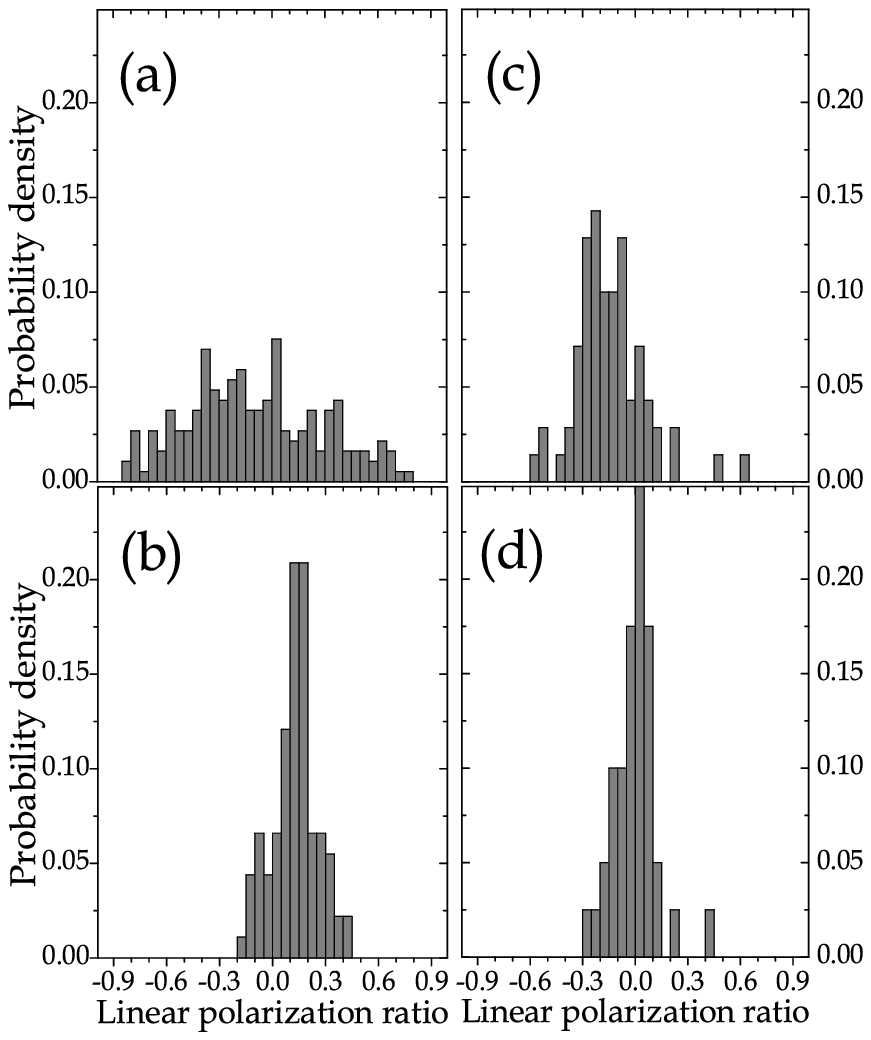}
\caption{Probability density of the linear polarization ratio in
single QDs. Dilute QD arrays : samples $A_{(II)}$ (a) and
$C_{(II)}$ (c). Dense QD arrays : samples $B_{(III)}$ (b) and
$C_{(III)}$ (d). $C$ was grown in a different laboratory from $A$
and $B$.} \label{Stat}
\end{figure}

To elucidate the condition for reaching, or reducing, the giant
optical anisotropy, we have performed systematic measurements of
the linear polarization ratio in roughly 400 QDs located in the
four samples described above. Similarly to QD1, we always find the
extrema of the PL signal intensity for the $X$ and $Y$ analysis
axes and the exchange splitting energy is smaller than 45 $\mu$eV.
We want here to highlight the absence of correlation between the
QD energy (ranging from 1.25 to 1.4 eV) and the linear
polarization ratio of the PL emission. In Fig.~\ref{Stat} we
display the histograms of the linear polarization ratio for
samples $A_{(II)}$ (a), $B_{(III)}$ (b), $C_{(II)}$ (c), and
$C_{(III)}$ (d). For each sample, the histogram is normalized to
the number of investigated QDs so that we get the probability
density of finding a given linear polarization ratio, in an
interval of 0.05 width.

The average linear polarization ratio $<$$P_L$$>$ ranges from
$<$$P_L$$>$=-0.007 in sample $C_{(III)}$ to -0.15 in sample
$C_{(II)}$ with $<$$P_L$$>$=0.12 and -0.09 in samples $B_{(III)}$
and $A_{(II)}$, respectively. The small values of $<$$P_L$$>$ show
that QD ensembles exhibit a weak polarization anisotropy in our
samples. Since there is no correlation between the QD energy and
the linear polarization ratio, identical values are obtained by
performing macroPL experiments in large mesas. These data are in
agreement with previous studies reporting a moderate optical
anisotropy in similar QD ensembles \cite{yu,cortez}.

With the single QD spectroscopy we gain new and useful information
by resolving the statistical distribution of the linear
polarization ratio. For samples $A_{(II)}$ and $B_{(III)}$ coming
from the same laboratory, we observe a strong increase of the
variance $\sigma_{P_L}$ when passing from the dense QD array
($\sigma_{P_L}$=0.12, Fig.~\ref{Stat}(b)) to the dilute QD array
($\sigma_{P_L}$=0.38, Fig.~\ref{Stat}(a)). For samples $C_{(II)}$
and $C_{(III)}$ coming from wafer $C$ grown in a \textit{distinct}
laboratory, we observe qualitatively the same behavior when
passing from a dense region ($\sigma_{P_L}$=0.11,
Fig.~\ref{Stat}(d)) to a dilute one ($\sigma_{P_L}$=0.2,
Fig.~\ref{Stat}(c)). As a consequence, we see that the giant
optical anisotropy is an intrinsic and universal characteristic of
the intermediate region (II), the so-called border region at the
onset of the Stranski-Krastanow growth mode where QD nucleation is
initiated. Despite the difference between samples grown in
separate laboratories, we see, in the prospect of future studies
in single QDs, that the border region can be exploited for
studying the giant optical anisotropy in a single quantum dot. On
the other hand, the possibility of restoring optical polarization
isotropy is optimal in dense QD arrays.

Recent studies of dilute QD arrays by high resolution near-field
imaging have revealed structural anisotropy as evidenced by QD
nucleation at the upper-step edge of terraces \cite{patella}.
Moreover, detailed AFM studies of the QDs nucleation and growth
\cite{gerardot} show a strongly anisotropic shape of the QDs as it
is covered by the capping GaAs. This shape anisotropy and
elongation along the [1$\overline{1}$0] axis is very pronounced in
the low QD density region of the sample. A microscopic model
taking into account the complex morphology and composition of the
QDs from dilute arrays is to be developed in order to explain the
very different oscillator strengths of the QD transitions as well
as the sign of $P_L$. In fact, the linear polarization ratio $P_L$
may be affected by the existence of an in-plane electric field, as
discussed in a forthcoming publication.

In conclusion we present the experimental evidence of giant
optical anisotropy in single InAs QDs which is attributed to a
strong difference of the oscillator strengths of the two bright
excitons. We show by systematic measurements that the giant
optical anisotropy is an intrinsic feature of dilute QD ensembles.

M. Terrier and S. Olivier are gratefully acknowledged for their
contributions to the processing of the mesa structures. LPA-ENS is
"unit\'{e} mixte (UMR 8551) de l'ENS, du CNRS, des Universit\'{e}s
Paris 6 et 7". This work is financially supported by the region
Ile de France through the project SESAME E-1751.

$^\ast$Electronic address: Guillaume.Cassabois@lpa.ens.fr

\newpage

\newpage
\begin{references}
\bibitem{li}X. Li, Y. Wu, D. Steel, D. Gammon, T. H. Stievater, D. S. Katzer, D. Park, C. Piermarocchi, and L. J. Sham, Science {\bf 301}, 809 (2003).
\bibitem{michler}P. Michler A. Kiraz, C. Becher, W. V. Schoenfeld, P. M. Petroff, L. Zhang, E. Hu, A. Imamoglu, Science {\bf 290}, 2282 (2000).
\bibitem{santori}C. Santori, D. Fattal, J. Vuckovic, G. S. Solomon, and Y. Yamamoto, Nature {\bf 419}, 594 (2002).
\bibitem{zrenner}A. Zrenner, E. Beham, S. Stufler, F. Findeis, M. Bichler, and G. Abstreiter, Nature {\bf 418}, 612 (2002).
\bibitem{benson}O. Benson, C. Santori, M. Pelton, and Y. Yamamoto, Phys. Rev. Lett. {\bf 84}, 2513 (2000).
\bibitem{pryor}C. E. Pryor and M. E. Flatt\'{e}, Phys. Rev. Lett. {\bf 91}, 257901 (2003).
\bibitem{zunger}G. Bester, S. Nair, and A. Zunger, Phys. Rev. B {\bf 67}, 161306 (2003).
\bibitem{ivchenko}E. L. Ivchenko, Phys. Status Solidi (a) {\bf 164}, 487 (1997).
\bibitem{toda}Y. Toda, S. Shinomori, K. Suzuki, and Y. Arakawa, Phys. Rev. B {\bf 58}, 10147 (1998).
\bibitem{bayer}M. Bayer, G. Ortner, O. Stern, A. Kuther, A. A. Gorbunov, A. Forchel, P. Hawrylak, S. Fafard, K. Hinzer, T. L. Reinecke, S. N. Walck, J. P. Reithmaier, F. Klopf, and F. Sch\"{a}fer, Phys. Rev. B {\bf 65}, 195315 (2002), and references therein.
\bibitem{finley}J. J. Finley, D. J. Mowbray, M. S. Skolnick, A. D. Ashmore, C. Baker, A. F. G. Monte, and M. Hopkinson, Phys. Rev. B {\bf 66}, 153316 (2002).
\bibitem{stevenson}R. M. Stevenson, R. M. Thompson, A. J. Shields, I. Farrer, B. E. Kardynal, D. A. Ritchie, and M. Pepper, Phys. Rev. B {\bf 66}, 081302 (2002).
\bibitem{urbaszek}B. Urbaszek, R. J. Warburton, K. Karrai, B. D. Gerardot, P. M. Petroff, and J. M. Garcia, Phys. Rev. Lett. {\bf 90}, 247403 (2003).
\bibitem{santoriB}C. Santori, D. Fattal, M. Pelton, G. S. Solomon, and Y. Yamamoto, Phys. Rev. B {\bf 66}, 045308 (2002).
\bibitem{kammerer}C. Kammerer, G. Cassabois, C. Voisin, M. Perrin, C. Delalande, Ph. Roussignol, and J. M. G\'{e}rard, Appl. Phys. Lett. {\bf 81}, 2737 (2002).
\bibitem{yu}P. Yu, W. Langbein, K. Leosson, J. M. Hvam, N. N. Ledentsov, D. Bimberg, V. M. Ustinov, A. Y. Egorov, A. E. Zhukov, A. F. Tsatsulnikov, and Y. G. Musikhin, Phys. Rev. B {\bf 60}, 16680 (1999).
\bibitem{henini}M. Henini, S. Sanguinetti, S. C. Fortina, E. Grilli, M. Guzzi, G. Panzarini, L. C. Andreani, M. D. Upward, P. Moriarty, P. H. Beton, and L. Eaves, Phys. Rev. B {\bf 57}, 6815 (1998).
\bibitem{cassabois}G. Cassabois, C. Kammerer, R. Sopracase, C. Voisin, C. Delalande, Ph. Roussignol, and J. M. G\'{e}rard, J. Appl. Phys. {\bf 91}, 5489 (2002).
\bibitem{marzin}J. Y. Marzin, J. M. G\'{e}rard, A. Izra\"{e}l, D. Barrier, and G. Bastard, Phys. Rev. Lett. {\bf 73}, 716 (1994).
\bibitem{uskov}A. V. Uskov, I. Magnusdottir, B. Tromborg, and J. M\o rk, and R. Lang, Appl. Phys. Lett. {\bf 79}, 1679 (2001).
\bibitem{cortez}S. Cortez, O. Krebs, P. Voisin, and J. M. G\'{e}rard, Phys. Rev. B {\bf 63}, 233306 (2001).
\bibitem{patella}F. Patella, S. Nufris, F. Arciprete, M. Fanfoni, E. Placidi, A. Sgarlata, and A. Balzarotti, Phys. Rev. B {\bf 67}, 205308 (2003).
\bibitem{gerardot}B. D. Gerardot, I. Shtrichman, D. Hebert, P. M. Petroff, J. Cryst. Growth, {\bf 252}, 44 (2003).
\end{references}
\end{document}